\documentclass[twocolumn,prb,amsmath,amssymb,showpacs,superscriptaddress,floatfix]{revtex4}

\usepackage{graphicx}

 \DeclareMathOperator{\Real}{\mathrm{Re}}
 \DeclareMathOperator{\sgn}{sgn}
 \DeclareMathOperator{\Tr}{Tr}

\begin{document}

 \title{Long-range odd triplet superconductivity in SF structures with N\'eel walls}

 \author{A.~F.~Volkov}
 \email{volkov@tp3.rub.de}
 \affiliation{Theoretische Physik III, Ruhr-Universit\"{a}t Bochum, D-44780 Bochum, Germany}
 \affiliation{Institute of Radioengineering and Electronics of the Russian Academy of Sciences, 103907 Moscow, Russia}

 \author{Ya.~V.~Fominov}
 \email{fominov@landau.ac.ru}
 \affiliation{L.~D.~Landau Institute for Theoretical Physics RAS, 119334 Moscow, Russia}

 \author{K.~B.~Efetov}
 \affiliation{Theoretische Physik III, Ruhr-Universit\"{a}t Bochum, D-44780 Bochum, Germany}
 \affiliation{L.~D.~Landau Institute for Theoretical Physics RAS, 119334 Moscow, Russia}

\date{27 October 2005}

\begin{abstract}
We consider a multidomain superconductor/ferromagnet (SF) structure with an in-plane magnetization, assuming that the
neighboring domains are separated by the N\'eel domain walls. We show that an odd triplet long-range component arises in
the domain walls and spreads into domains over a long distance of the order $\xi_T = \sqrt{D/2\pi T}$ (in the dirty
limit). The density of states variation in the domains due to this component changes over distances of the order $\xi_T$
and turns to zero in the middle of domains if the magnetization rotates in the same direction in all domain walls.
\end{abstract}


% 74.45.+c Proximity effects; Andreev effect; SN and SNS junctions

% 74.78.-w Superconducting films and low-dimensional structures
% 74.78.Fk Multilayers, superlattices, heterostructures

% 74.50.+r Tunneling phenomena; point contacts, weak links, Josephson effects (for SQUIDs, see 85.25.Dq; for Josephson
% devices, see 85.25.Cp; for Josephson junction arrays, see 74.81.Fa)

% 75.70.-i Magnetic properties of thin films, surfaces, and interfaces (for magnetic properties of nanostructures, see
% 75.75.+a)
% 75.70.Kw Domain structure (including magnetic bubbles)

\maketitle

\section{Introduction}

The past decade has seen a rapid growth of interest in the study of hybrid superconductor-ferromagnet (SF) structures
(see, for example, reviews \onlinecite{Buzdin,LP,BVE_review}). The interest in such systems originates from the
possibility of finding new physical phenomena as well from the hope of constructing new devices based on these
structures. New physical phenomena arising in these systems are the result of a nontrivial interplay of competing types
of ordering in superconductors and ferromagnets. Superconducting correlations lead in superconductors to the appearance
of the Cooper pairs, that is, the pairs of electrons with opposite spins. On the opposite, the exchange interaction in
ferromagnets tries to align the electron spins in one direction. In SF structures these two types of interactions are
spatially separated and can coexist despite much greater value of the exchange energy $h$ in comparison to the
superconducting order parameter $\Delta$.

Due to the proximity effect \cite{deGennes} the superconducting correlations penetrate the ferromagnet in SF structures.
The opposite effect, i.e., the penetration of a magnetic moment $\mathbf M$ into the superconductor, also takes place.
It turns out that the magnetic moment $\mathbf M_S$ is induced in the superconductor. The magnetic moment $\mathbf M_S$
is aligned in the direction opposite to the magnetization direction of free electrons in the ferromagnet and spreads
over a distance of the order of the superconducting correlation length $\xi_S =\sqrt{D_S /\Delta}$ (in the dirty limit).
\cite{BVEscr} On the other hand the condensate wave function $f$ penetrates the ferromagnet with an uniform
magnetization $\mathbf M_F$ over a much shorter distance of the order of the ``exchange length'' $\xi_h
=\sqrt{D_F/h}$.\cite{Buzdin,LP} The condensate wave function decays in F in a nonmonotonic way as $f(x) \sim \exp
(-x/\xi_h)\cos (x/\xi_h)$; it oscillates in space and decreases exponentially. This nonmonotonic behavior of $f(x)$
leads to a nonmonotonic dependence of the critical temperature $T_c$ of the superconducting transition in SF bilayers
and multilayers\cite{Buzdin,LP,Jiang,Muehge,Aarts,Lazar,Garifullin} and to a $\pi$-state in SFS Josephson
junctions.\cite{Buzdin,LP,Golubov,Ryazanov,Kontos,Blum,Strunk,Sellier}

In the case of a nonuniform magnetization in the ferromagnet a new phenomenon appears: a triplet component of the
condensate wave function $f$ (generally speaking the condensate wave function is a matrix in the particle-hole and spin
space) arises in the SF system.\cite{BVE_review} This triplet component is an odd function of the Matsubara frequency
$\omega$ (while the conventional BCS singlet component of $f$ is an even function of $\omega$) and spreads in the
ferromagnet over a long distance of the order of $\xi_T =\sqrt{D_F /2\pi T}$. This long-range triplet odd-frequency
component was predicted to exist in a SF structure with nonhomogeneous magnetization in Ref.~\onlinecite{BVElong} and
this prediction was confirmed for a slightly different case in Ref.~\onlinecite{Kadigrobov}. In
Refs.~\onlinecite{BVElong} and~\onlinecite{Kadigrobov}, a SF structure with a domain wall at the SF interface was
considered, that is, it was assumed that $\mathbf M_F =M \bigl( 0,\sin\alpha (x), \cos\alpha (x) \bigr)$, where $\alpha
(x)=Qx$ in the interval $0<x<a_Q$ and $\alpha (x)= Q a_Q$ at $x>a_Q$. The triplet component $f_L$ was shown to arise in
the domain wall and to penetrate the ferromagnet over a long distance $\xi_T$. Unlike the triplet component in
superfluid $^3$He and in Sr$_2$RuO$_4$, this odd triplet component corresponds to \textit{s}-wave correlations and hence
is symmetric in the momentum space; therefore it is not destroyed by scattering on ordinary, nonmagnetic impurities, and
survives in the dirty limit. We call this component the long-range triplet component (LRTC). The LRTC may also arise in
a SF structure with a uniform magnetization and spin-active interface.\cite{Eschrig1} Note that from a macroscopic point
of view a domain wall at the SF interface can also be considered as a ``spin-active interface''. The LRTC may arise in a
multilayered SF structure with noncollinear orientations of the magnetization vector $\mathbf M_{F_i}$ in different
F$_i$ layers.\cite{VBE,BVEmanif} In particular, a new type of superconductivity (odd triplet superconductivity) has been
predicted in such structures if the thickness of the F layers $d$ obeys the condition: $\xi_h \ll d \lesssim \xi_T$. In
this case the Josephson coupling between neighboring F layers is realized only via the LRTC because the singlet
component decays very fast in the F layers. Therefore superconductivity in the transverse direction is due to the LRTC,
whereas in-plane superconductivity is caused mainly by the singlet BCS component. The influence of the LRTC on the
critical temperature of the superconducting transition in FSF structures with a noncollinear magnetization orientation
was studied in Ref.~\onlinecite{Fominov}.

Historically, the odd-frequency triplet pairing was conjectured in 1974 by Berezinsii\cite{Berez} as a possible
mechanism for superfluidity in $^3$He. It turned out later that in $^3$He another type of triplet pairing (even in
frequency, odd in momentum) is realized. Odd-frequency pairing in solids was also studied in
Refs.~\onlinecite{B+K,C+T,Balatsky}. A triplet odd-frequency pairing was investigated in Refs.~\onlinecite{B+K}
(two-dimensional electron gas with repulsion in the presence of impurities) and \onlinecite{C+T} (a Kondo lattice
model). A singlet odd-frequency pairing was analyzed\cite{Balatsky} as a possible type of pairing in high-temperature
superconductors.

Although several experimental results may be interpreted in terms of the
LRTC,\cite{Petrashov1,Giordano,Giroud,Petrashov2,Pena,Petrashov3} there are still no direct experimental evidences in
favor of the odd triplet superconductivity. Therefore there is a need to investigate, in more detail, a possibility to
observe this new type of superconductivity. One of the important and inherent features of ferromagnets is the domain
structure. The domain structure in a ferromagnet may essentially alter the properties of SF structures. For example,
modulation of the phase difference $\varphi$ between two superconductors due to internal magnetic fields in the
ferromagnetic domains, can lead to a negative critical current $I_c$ in the Josephson SFS junction.\cite{VA} The
influence of domains on $I_c$ in SFS Josephson junctions was studied in Refs.~\onlinecite{BVEcrcur}
and~\onlinecite{Blanter}. In Ref.~\onlinecite{BVEcrcur} the magnetization vector is assumed to be in-plane and to rotate
around the direction normal to the plane of the junction. In this case the LRTC arises, and the possible $\pi$-state may
be suppressed due to an effective averaging of the exchange field. In Ref.~\onlinecite{Blanter} a SFS junction with two
in-plane domains of opposite orientation was studied (no LRTC arises in this case). It was shown that if the thicknesses
of the domains are equal, the critical current $I_c$ is always positive.

In realistic domain structures, the domains are separated by domain walls. Below we shall discuss the case when the
magnetization vector $\mathbf M_F$ lies in the plane of the F film and the domain walls are of the N\'eel
type.\cite{Aharoni} A limiting case of such a structure was considered in Refs.~\onlinecite{BEL}
and~\onlinecite{Eschrig2}. It was assumed that the vector $M_F$ rotates in space continuously
\begin{equation} \label{M(y)}
\mathbf M_F (y) = M_0 (0,\sin Qy,\cos Qy)
\end{equation}
(we choose the $x$ axis normal to the plane of the F film, whereas in Ref.~\onlinecite{Eschrig2} the $z$ axis is normal
to the plane of the F film). The possibility of a cryptoferromagnetic state in SF structures with the magnetization
$M_F(y)$ given by Eq. (\ref{M(y)}) was studied in Ref.~\onlinecite{BEL}. The F layer was supposed to be very thin: $d
\ll \xi_h$. A solution for the Eilenberger equation has been found near the critical temperature $T_c$ of the
superconducting transition. It was established that in a certain interval of parameters ($Q$, $d$, etc.) a homogeneous
state in the ferromagnet becomes energetically unfavorable and the nonhomogeneous magnetization determined by Eq.
(\ref{M(y)}) arises in the F film. In Ref.~\onlinecite{Eschrig2} a SF bilayer with the F film of arbitrary thickness was
studied in the dirty limit. Analysis of a solution for the Usadel equation shows that in the case of magnetization,
uniformly rotating along the F film [see Eq. (\ref{M(y)})], the condensate function $f$ penetrates into the F film over
a short distance of the order $\xi_h$. The LRTC is absent in this case. According to
Refs.~\onlinecite{BVElong,Kadigrobov,Eschrig1,BVEmanif}, the LRTC appears if $\mathbf M_F$ rotates across the F layer
(i.e., depends not on $y$ but on $x$).

\begin{figure}
 \includegraphics[width=7cm]{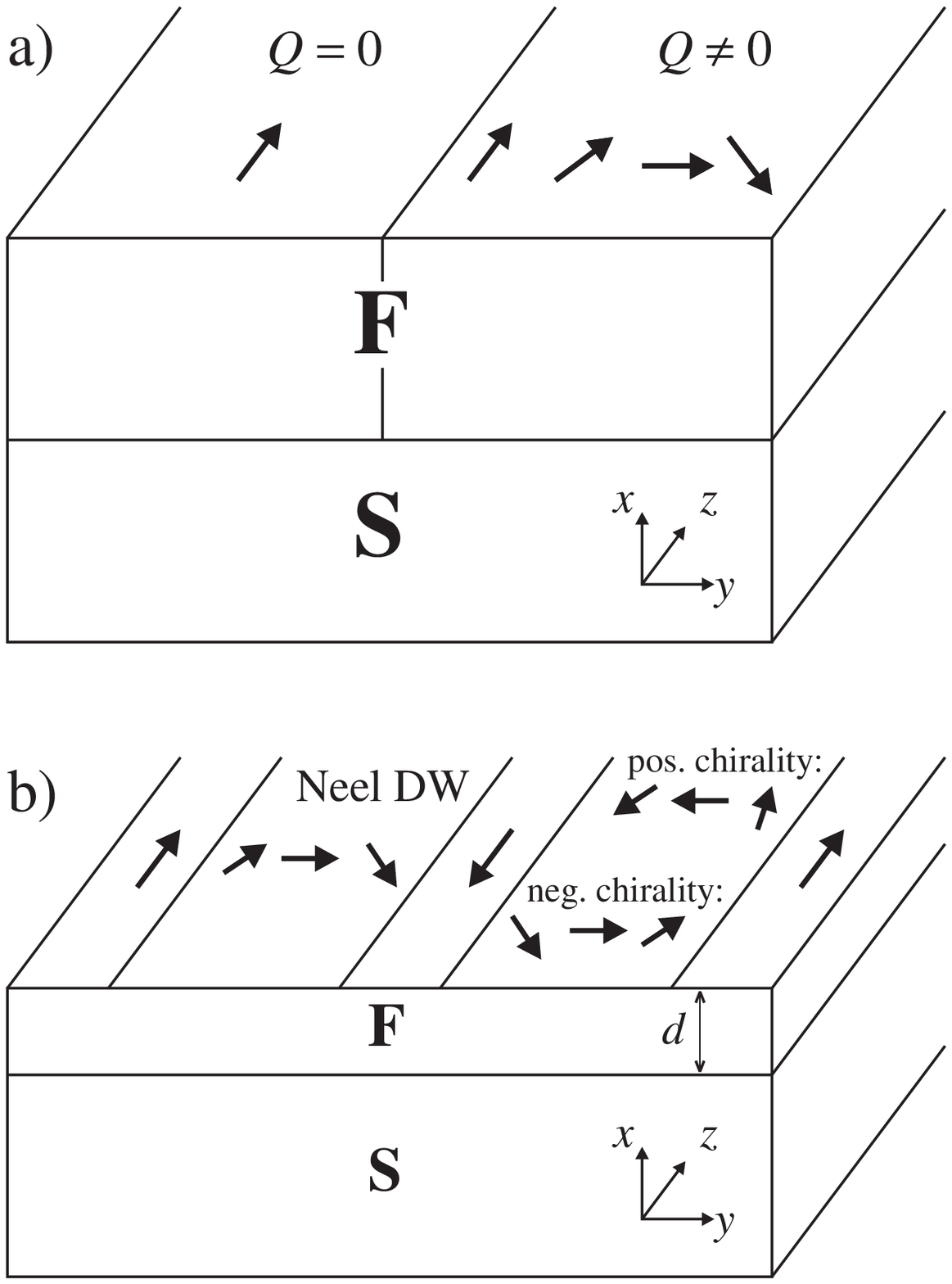}
\caption{SF systems considered in the paper. (a)~SF bilayer with half-infinite S and F parts. The domain ($y<0$) and the
region with rotating magnetization ($y>0$) are also half-infinite. (b)~Multidomain F layer of thickness $d$ in contact
with a bulk superconductor. Depending on the relative orientation of rotating magnetizations in neighboring domain
walls, we distinguish the cases of positive and negative chirality ($Q$ has the same or opposite sign in the neighboring
domain walls, respectively). The proportion between the widths of domains and domain walls is chosen only for drawing
purposes.}
 \label{fig:system}
\end{figure}

In the present paper we consider a domain structure in a thin F film, where domains with antiparallel in-plane
magnetizations are separated by the N\'eel walls (while the magnetization does not change across the thin F film). This
domain structure is realized in real ferromagnetic films.\cite{Aharoni} The $yz$ plane is chosen to be parallel to the
SF interface (see Fig.~\ref{fig:system}). We show that the LRTC arises at the N\'eel domain walls and decays
exponentially away from the domain walls and the SF interface over a long distance $\xi_T$. We calculate the density of
states (DOS) variation $\delta\nu$ in the ferromagnet caused by the proximity effect and find that $\delta \nu (y)$
turns to zero in the middle of domains in the case of positive chirality.

The paper is organized as follows. In Sec.~\ref{sec:basic}, we formulate the Usadel equations, the corresponding
boundary conditions, and investigate the main features of the long-range triplet superconducting component which appears
due to the presence of N\'eel domain walls. The analysis is made on the simplified model with half-infinite S and F
layers and only one half-infinite region with rotating magnetization. In Sec.~\ref{sec:inf}, we consider the case of the
multidomain F layer, employing the results of Sec.~\ref{sec:basic}. To make the model realistic, in Sec.~\ref{sec:multd}
we take into account a finite thickness of the ferromagnet. To illustrate the results for the LRTC, we study the density
of states due to it. Finally, we present our conclusions in Sec.~\ref{sec:concl}.

\section{Basic equations. Half-infinite domain} \label{sec:basic}

Consider a ferromagnet ($x>0$) in contact with a superconductor ($x<0$). We assume that the in-plane exchange field
$\mathbf h(y)$ in the F layer, which is proportional to the magnetization $\mathbf M_F$, depends on $y$: $\mathbf h (y)
= h (0,0,1)$ at $y<0$ and $\mathbf h (y)=h (0,\sin \alpha (y),\cos \alpha (y))$ with $\alpha (y)=Qy $ at $y>0$. This
means that the magnetization vector $\mathbf M_F$ is oriented along the $z$ axis at $y<0$ and rotates in the $yz$ plane
at $y>0$. The region with rotating magnetization models a N\'eel domain wall. The structure that we discuss first is
shown in Fig.~\ref{fig:system}(a) and contains only one half-infinite domain and one half-infinite region with rotating
magnetization; the thickness of the ferromagnet is also infinite. Then we shall use the obtained results to describe a
realistic structure depicted in Fig.~\ref{fig:system}(b).

Our goal is to find the condensate Green functions in the ferromagnet, induced due to the proximity effect. We consider
the dirty limit, which means, in particular, that $h\tau \ll 1$, where $\tau$ is the momentum relaxation time due to
elastic scattering.

We employ a widely used model: an exchange field $h$ acts on free electrons and there is no attractive interaction in
the F layer leading to the superconducting order parameter.\cite{Buzdin,LP,BVE_review} However, the condensate (Gor'kov)
functions are finite in the F region due to the boundary conditions at the SF interface. We are interested in distances
much larger than the Fermi wavelength, therefore we can use the quasiclassical Green functions $\check g^{R(A)}$. The
matrix retarded (advanced) functions $\check g^{R(A)}$ in the ferromagnet obey the Usadel equation\cite{BVE_review}
\begin{equation} \label{usadel}
D\nabla \left( \check g^{R(A)} \nabla \check g^{R(A)} \right) +i \varepsilon \left[ \hat\tau_3, \check g^{R(A)} \right]-
i \left[ \check h, \check g^{R(A)} \right] =0,
\end{equation}
where $\varepsilon$ is energy, $\check h = h \bigl( \hat\tau_3 \hat\sigma_3 \cos \alpha (y)+\hat\tau_0 \hat\sigma_2 \sin
\alpha (y) \bigr)$, and $\alpha (y)=0$ at $y\leqslant 0$ while $\alpha (y)=Qy$ at $y\geqslant 0$. We assume that the
diffusion constants $D$ for electrons with spin up and down are equal (this is correct if the exchange energy $h$ is
much less than the Fermi energy $\varepsilon_F$). The matrix Green functions $\check g$ are $4\times 4$ matrices in the
Gor'kov-Nambu and spin spaces. $\hat\tau_i$ and $\hat\sigma_i$ are the Pauli matrices in the Gor'kov-Nambu and spin
spaces, respectively.

We represent $\check g^{R(A)}$ in the form
\begin{equation}
\check g^{R(A)} = \pm \hat\tau_3 \hat\sigma_0 +\check f ,
\end{equation}
where the first term is the retarded (advanced) quasiclassical Green function in the normal state. The superconducting
correlations, described by $\check f(x,y)$, are assumed to be weak due to the finite interface transparency (resulting
from an oxide barrier or from mismatch in the Fermi surfaces of S and F materials). In the considered case of weak
proximity effect ($|\check{f}| \ll 1$), Eq. (\ref{usadel}) can be linearized. In the Matsubara representation it
acquires the form
\begin{multline} \label{usadelM}
\nabla^2 \check f - 2 k_\omega^2 \check f - i k_h^2 \sgn\omega \Bigl( \left\{ \hat\sigma_3, \check f
\right\} \cos\alpha (y) \\
+\hat\tau_3 \left[ \hat\sigma_2, \check f \right] \sin\alpha (y) \Bigr) =0,
\end{multline}
where $\omega = \pi T (2n+1)$, $k_\omega^2 = |\omega |/D$, $k_h^2 = h /D$, the square brackets denote the commutator,
and the braces denote the anticommutator.

The Green function in the bulk of the superconductor is $\check f_S = \hat\tau_2 \hat\sigma_3 f_S$, with $f_S =\Delta /
\sqrt{\omega^2 +\Delta^2}$. We shall use the following boundary condition for $\check f$ at the SF interface ($x=0$):
\begin{equation} \label{BC}
\frac{\partial \check f}{\partial x} =-\frac{\check f_S}{\gamma_b} ,
\end{equation}
where $\gamma_b = R_b \sigma$, while $\sigma$ is the conductivity of the ferromagnet and $R_b$ is the interface
resistance per unit area. This boundary condition follows from the general ones\cite{Zaitsev,Kupriyanov} if two
assumptions are made: (1) the proximity effect is weak (i.e., $\gamma_b /\xi_h \gg 1$) and (2) the bulk solution $\check
f_S$ in the superconductor is unperturbed and valid up to the interface (i.e., $\gamma_b /\xi_S \gg \sigma/\sigma_S$).

We can rewrite Eq. (\ref{usadelM}) in the form taking into account boundary condition (\ref{BC}):
\begin{multline} \label{usadelM1}
\frac{\partial^2 \check f}{\partial x^2} +\frac{\partial^2 \check f}{\partial y^2} - 2 k_\omega^2 \check f - i k_h^2
\sgn\omega \Bigl( \left\{ \hat\sigma_3, \check f \right\} \cos\alpha (y) \\
+\hat\tau_3 \left[ \hat\sigma_2, \check f \right] \sin\alpha (y) \Bigr) = -\frac{2 \check f_S}{\gamma_b} \delta (x) .
\end{multline}
We must seek for an even solution of this equation, then this is equivalent to the problem with the boundary condition
(we have reflected $\check f$ with respect to $x=0$ and now solve the equation at all $x$). However, the requirement
that the solution is even, will be automatically satisfied: as we shall see below, the Fourier harmonics (over $x$)
depend only on $k^2$, hence they are even in $k$, which means that $f(x)$ is even in $x$.

Performing the Fourier transformation $\check f(k,y) =\int dx\check f(x,y) \exp(-ikx)$, we obtain
\begin{multline} \label{usadelM2}
\frac{\partial^2 \check f}{\partial y^2} -(k^2+ 2 k_\omega^2) \check f - i k_h^2 \sgn\omega \Bigl( \left\{ \hat\sigma_3,
\check f \right\} \cos\alpha (y) \\
+\hat\tau_3 \left[ \hat\sigma_2, \check f \right] \sin\alpha (y) \Bigr) =-\frac{2 \check f_S}{\gamma_b}.
\end{multline}

At $y>0$ the function $\alpha (y)$ is $y$-dependent, while at $y<0$ we have $\alpha =0$. In the region of positive $y$
one can exclude the $y$-dependence from Eq. (\ref{usadelM2}) with the aid of rotation
\begin{equation} \label{Utrans}
\check f =\check U \check f_u \check U^+,
\end{equation}
where $\check U =\exp \bigl( i\hat\tau_3 \hat\sigma_1 \alpha (y)/2 \bigr)$. As a result, we get ($y>0$)
\begin{multline} \label{usadelM3}
\frac{\partial^2 \check f_u}{\partial y^2}- \left( k^2+\frac{Q^2}2 + 2 k_\omega^2 \right) \check f_u -\frac{Q^2}2
\hat\sigma_1 \check f_u \hat\sigma_1 \\
+i Q \hat\tau_3 \bigl\{ \hat\sigma_1, \frac{\partial \check f_u}{\partial y} \bigr\} -i k_h^2 \sgn\omega \bigl\{
\hat\sigma_3, \check f_u \bigr\} = -\frac{2 \check f_S}{\gamma_b}
\end{multline}
in terms of the new function $\check f_u(k,y)$. The same equation is valid for $y<0$ if we set $Q=0$:
\begin{equation} \label{usadelM4}
\frac{\partial^2 \check f_u}{\partial y^2} -\left( k^2 + 2 k_\omega^2 \right) \check f_u -i k_h^2 \sgn\omega \left\{
\hat\sigma_3, \check f_u \right\}  =-\frac{2 \check f_S}{\gamma_b}.
\end{equation}

The original functions $\check f$ and $\partial \check f /\partial y$ are continuous at $y=0$. Therefore the rotated
functions obey the following boundary conditions at $y=0$:
\begin{align}
\check f_u (-0) &= \check f_u (+0),  \label{MC1} \\
\frac{\partial \check f_u (-0)}{\partial y} &= \frac{\partial \check f_u (+0)}{\partial y} +i \frac Q2 \hat\tau_3
\left\{ \hat\sigma_1 , \check f_u \right\} . \label{MC2}
\end{align}

Thus we have to solve the linear matrix Eqs. (\ref{usadelM3}) ($y>0$) and (\ref{usadelM4}) ($y<0$) of the second order
with the boundary conditions (\ref{MC1}) and (\ref{MC2}) at $y=0$. We can represent the solution in the form
\begin{equation} \label{f(y)}
\check f_u =\check F(Q) \theta (y)+\check F(0)\theta (-y)+\delta \check f_u,
\end{equation}
where $\theta$ is the Heaviside step function and the constants $\check F(Q)$ and $\check F(0)$ are the homogeneous
solutions of Eqs. (\ref{usadelM3}) and (\ref{usadelM4}) at $y=\pm\infty$. The matrices $\check F$ have the form
\begin{equation} \label{F}
\check F = \hat\tau_2 ( \hat\sigma_0 F_0 +\hat\sigma_3 F_3 ),
\end{equation}
where
\begin{align}
F_0 (Q) &= -\frac{4i f_S k_h^2 \sgn\omega}{\gamma_b \mathcal D(Q)},  \label{F_0} \\
F_3 (Q) &= \frac{2 f_S \left( k^2+Q^2+2 k_\omega^2 \right)}{\gamma_b \mathcal D(Q)}, \label{F_3}
\end{align}
and
\begin{equation}
\mathcal D(Q)= \left( k^2+Q^2+2 k_\omega^2 \right) \left( k^2+2 k_\omega^2 \right)+4 k_h^4.
\end{equation}

The correction $\delta \check f_u(k,y)$ obeys the same Eqs. (\ref{usadelM3}) and (\ref{usadelM4}) without the right-hand
side. It has the form
\begin{equation} \label{delta-f}
\delta \check f_u = \hat\tau_2 \hat\sigma_3 f_3 + \hat\tau_2 \hat\sigma_0 f_0 + \hat\tau_1 \hat\sigma_1 f_1.
\end{equation}
The first term is the singlet component. The second term is the triplet component with zero projection of the Cooper
pair spin on the $z$ axis. This component arises even in the case of a homogenous magnetization of the ferromagnet and
decays in the F film over the short distance $\xi_h$. The last term in Eq. (\ref{delta-f}) is the triplet component with
the spin moment projection $\pm 1$. It arises in the case of a nonhomogeneous magnetization and decays over a long
distance of the order $\xi_T$. The functions $f_i(k,y)$ in Eq. (\ref{delta-f}) can be represented as a sum of
eigenfunctions of Eqs. (\ref{usadelM3}) and (\ref{usadelM4}), i.e.,
\begin{align}
f_i (y) &= \sum_l A_{il} \exp \bigl( -\kappa_l (Q) y \bigr),\quad \text{at } y>0  \label{A}, \\
f_i (y) &= \sum_l B_{il} \exp \bigl(  \kappa_l (0) y \bigr),\quad \text{at } y<0  \label{B}.
\end{align}
The inverse decay lengths $\kappa_l(Q)$ are the eigenvalues of Eqs. (\ref {usadelM3}) and (\ref{usadelM4}) (without the
right-hand side). The equation for $\kappa_l(Q)$ has the form ($l=1,2,3$)
\begin{multline}
\left[ \left( \kappa_l^2 -k^2 -Q^2 -2 k_\omega^2 \right)^2 + 4 \left( Q\kappa_l \right)^2 \right] \left( \kappa_l^2 -k^2
-2 k_\omega^2 \right) \\
+4 k_h^4 \left( \kappa_l^2 -k^2 -Q^2 -2 k_\omega^2 \right) = 0.
\end{multline}
We assume that the exchange length is the shortest length in the problem:
\begin{equation} \label{Assum}
k_h^2 \gg k^2, Q^2, k_\omega^2 .
\end{equation}
Then the eigenvalues $\kappa_l$ consist of two ``short-range'' values
\begin{equation}
\kappa_\pm \approx \left( 1\mp i \sgn\omega \right) k_h,
\end{equation}
and one ``long-range'' value
\begin{equation}
\kappa_L (Q) \approx \sqrt{k^2+Q^2+2 k_\omega^2}.
\end{equation}
At $y<0$ we have the same $\kappa_l$ with $Q=0$.

Calculating the corresponding eigenvectors under assumption (\ref{Assum}), in the first order over $1/k_h$ we obtain
\begin{equation}
A_{0\pm}\approx \mp A_{3\pm},\quad A_{1\pm}\approx \mp\frac{2Q}{\kappa_\pm} A_{3\pm},\quad A_{3,0L}\approx 0 .
\end{equation}
To be more exact, $A_{3L} \approx \left( Q\kappa_L /i k_h^2 \sgn\omega \right) A_{1L}$ and $A_{0L}$ is even smaller. The
same relations with $Q=0$ hold for $B_{il}$, which yields
\begin{equation}
B_{0\pm}\approx \mp B_{3\pm}
\end{equation}
for nonzero coefficients.

The next step is to match solutions (\ref{A}) and (\ref{B}) with the help of boundary conditions (\ref{MC1}) and
(\ref{MC2}) and to find the coefficients $A_{il}$ and $B_{il}$. This simple but cumbersome calculation is presented in
the Appendix. In the considered limit of a small exchange length [see Eq. (\ref{Assum})], the coefficients
$A_{1L}\approx B_{1L}$ [see Eq. (\ref{A1LB1L})] that describe the LRTC are the largest ones. In this limit the function
$F_0 (Q)$ has a simple form (\ref{F0}). Therefore the magnitude of the LRTC at the interface between a domain and a
domain wall ($y=0$) is equal to
\begin{equation} \label{f_Ltr}
f_L (k,0) \equiv f_1 (k,0) =-\frac{i f_S \sgn\omega}{\gamma_b k_h^2} \frac Q{\kappa_Q +\kappa_0},
\end{equation}
where for brevity we have denoted
\begin{equation}
\kappa_Q \equiv \kappa_L(Q),\quad \kappa_0 \equiv \kappa_L(0).
\end{equation}

Now we return to the real space and analyze our results. The spatial dependence of the LRTC at $y<0$ is [see Eq.
(\ref{B})]
\begin{equation} \label{f(x,y)}
f_L (x,y)=\int \frac{dk}{2\pi} f_L (k,0) \exp (ikx) \exp \bigl( \kappa_0 y \bigr) .
\end{equation}
From this formula one can easily find the asymptotic behavior of the LRTC $f_L(x,y)$. For large negative $y$ ($|y| \gg
1/ k_\omega$), we expand $\kappa_0$ and $\kappa_Q$ with respect to $(k/ k_\omega)^2$ (since the characteristic $k$ in
the integral is of the order $\sqrt{k_\omega /|y|}$) and obtain
\begin{multline} \label{f(x,y)As}
f_L (x,y) =-\frac{i f_S Q \sgn\omega}{\gamma_b k_h^2} \sqrt{\frac{k_\omega}{\sqrt 2 \pi |y|}} \\
\times \frac{\exp \left( -\sqrt 2 k_\omega |y| \right) \exp \left( - x^2 k_\omega /\sqrt 2 |y| \right)}{\sqrt 2 k_\omega
+\sqrt{2 k_\omega^2 +Q^2}}.
\end{multline}
This formula shows that the condensate function $f_L (x,y)$ decays exponentially with increasing $y$, but the
characteristic length is rather large ($\sim k_\omega^{-1}$).

For comparison, we can calculate the short-range component at the SF interface in the case of a homogenous
magnetization: $f_0 = -i f_S \sgn\omega / 2 \gamma_b k_h$ [it follows directly from Eq. (\ref{usadelM4})]. Formula
(\ref{f(x,y)As}) shows that at the SF interface ($x=0$) and at distances $|y| \lesssim k_\omega^{-1}$, the function $f_L
(x,y)$ is of the order $f_S \min( Q, k_\omega) /\gamma_b k_h^2$, which is smaller than the amplitude of the short-range
component by the parameter $\min( Q, k_\omega) /k_h$. Thus, the interface amplitude of the LRTC is smaller, however it
decays much slower in space.

In the domain wall ($y>0$) the behavior of the function $f_L (x,y)$ is nearly the same as at $y<0$ [Eq.
(\ref{f(x,y)As})] if $Q < k_\omega$. In the opposite limit $Q > k_\omega$ the function $f_L (x,y)$ in the domain wall
decays faster: $f_L (x,y) \propto \exp (-Q|y|)$.

Having found the condensate function $\check f$, we can calculate the density of states (DOS) in the ferromagnetic
region. The DOS, normalized to the normal-metallic value, is given by the general formula
\begin{equation}
\nu (\varepsilon)= \left. \frac 14 \Real \Tr \left( \hat\tau_3 \hat\sigma_0 \check g \right) \right|_{\omega \rightarrow
-i\varepsilon} .
\end{equation}
Using the normalization condition $\check g^2 =1$ and the smallness of the condensate function, we can write the
correction to the DOS due to the proximity effect as
\begin{equation} \label{DOSa}
\delta \nu (\varepsilon) = - \left. \frac{\Real f_L^2}2 \right|_{\omega \rightarrow -i\varepsilon}
\end{equation}
(we consider the region in space where only the LRTC is essential). This expression is valid at any $Q$, both zero and
nonzero. Equations (\ref{f(x,y)As}) and (\ref{DOSa}) show that the LRTC changes the DOS in the ferromagnet at distances
much larger than the exchange length $\xi_h$.

\section{Triplet component in multidomain SF structures} \label{sec:inf}

In this section we study the LRTC in a SF structure with a multidomain ferromagnetic layer, still assuming infinite
thickness of the F layer [Fig.~\ref{fig:system}(b) with $d\to\infty$]. One can distinguish between two possibilities:
(a)~positive chirality, when the magnetization vector $\mathbf M(y)$ in all the domain walls rotates in the same
direction (e.g., clockwise), and (b)~negative chirality, when the vector $\mathbf M(y)$ in neighboring domain walls
rotates in the opposite directions [e.g., clockwise in the $2n$-th domain walls and counterclockwise in the $(2n+1)$-th
domain walls]. We are interested in the LRTC assuming that the exchange length $\xi_h$ is much smaller than the
coherence length $\xi_T$. At distances $x$ essentially exceeding the length $\xi_h$ only the LRTC survives in the F
layer.

We assume that the width of the domains with $Q=0$ is $2 a_0$ and the width of the domain walls ($Q\neq 0$) is $2 a_Q$.
The origin ($y=0$) is located in the middle of a domain with the constant magnetization. At $x \gg \xi_h$ only the
long-range components of the condensate function survive in the ferromagnet. The largest long-range component is the
LRTC. At the boundary between a domain and a domain wall the solution must satisfy boundary conditions (\ref{MC1}) and
(\ref{MC2}). Consider first the case of positive chirality. The angle $\alpha(y)$ is then an odd function of $y$, which
means that $f_1(y)$ is also odd --- this general symmetry can be demonstrated in Eq. (\ref{usadelM1}). Hence the
solution for the LRTC is
\begin{align}
f_1 (y) &= A \sinh (\kappa_0 y),\quad -a_0 < y<a_0, \\
f_1 (y) &= B \sinh \bigl( \kappa_Q (y-a_0-a_Q) \bigr),\quad a_0< y < a_0+2a_Q .
\end{align}
Matching these solutions and their derivatives at $y= a_0$, we find
\begin{equation}
B = -A \frac{\sinh \theta_0}{\sinh \theta_Q} =- \frac{Q F_0}{\cosh \theta_Q \left( \kappa_Q +\kappa_0 \frac{\tanh
\theta_Q}{\tanh \theta_0} \right)},
\end{equation}
where $\theta_Q =\kappa_Q a_Q$ and $\theta_0 =\kappa_0 a_0$. The amplitude of the LRTC at $y=a_0$ is
\begin{equation} \label{f1a0}
f_1 (a_0) =\frac{Q F_0}{\kappa_Q \coth \theta_Q +\kappa_0 \coth \theta_0}.
\end{equation}

We see that $f_1 (a_0)$ turns to zero both at $a_Q \to 0$ and $a_0 \to 0$. These limits mean that the widths of the
domain walls and domains are assumed to be small in comparison with $\xi_T$ while larger than $\xi_h$. The case $a_Q =0$
implies that we have a domain structure with collinear magnetization orientation. The case $a_0 =0$ corresponds to a SF
structure with continuously rotating magnetization (the case studied in Ref.~\onlinecite{Eschrig2}). In both cases, the
LRTC does not arise.

The spatial dependence of the LRTC in the domain ($|y|<a_0$), corresponding to Eq. (\ref{f1a0}), is given by the inverse
Fourier transformation
\begin{equation} \label{f_1Dom}
f_L (x,y) =\int \frac{dk}{2\pi} e^{ikx} f_1(a_0) \frac{\sinh (\kappa_0 y)}{\sinh \theta_0} .
\end{equation}
Interestingly, the function $f_L(x,y)$ turns to zero in the center of a domain ($y=0$). This means that the DOS
variation due to the LRTC also turns to zero in the domain center.

Consider now the case of negative chirality, when the $\mathbf M$ vector rotates in the opposite directions in
neighboring domain walls. In this case the spatial dependence of the function $f_1 (y)$ in domain walls remains the same
as before, i.e., this function is an odd function with respect to the center of a domain wall. However the spatial
dependence of the LRTC in domains changes drastically: it becomes an even function with respect to the center of a
domain. Therefore this dependence is
\begin{align}
f_1 (y) &= C \cosh (\kappa_0 y),\quad -a_0 < y < a_0, \\
f_1 (y) &= D \sinh \bigl( \kappa_Q (y-a_0-a_Q) \bigr),\quad a_0 < y< a_0+2a_Q.
\end{align}
From boundary conditions (\ref{MC1}) and (\ref{MC2}) we find the coefficients $C$ and $D$, and finally
\begin{equation} \label{f1a0_nc}
f_1 (a_0) =\frac{Q F_0}{\kappa_Q \coth \theta_Q +\kappa_0 \tanh \theta_0}.
\end{equation}

In this case the LRTC disappears only in the limit $a_Q \to 0$ because in this limit one again has a domain structure
with collinear magnetization orientation and very narrow domain walls.

Another type of SF structures, sensitive to the chirality of the vector $\mathbf M$, was considered in
Refs.~\onlinecite{VBE} and~\onlinecite{BVEmanif}. It was shown that the sign of the critical Josephson current in a
multilayered SF structure depends on chirality.

\section{Finite thickness of multidomain F layer} \label{sec:multd}

In this section we consider a realistic structure with a ferromagnetic layer of finite thickness $d$, see
Fig.~\ref{fig:system}(b). We again have to solve Eq. (\ref{usadelM}) with boundary conditions. The first of them, at the
SF interface, is Eq. (\ref{BC}) and the other one, at the free surface of the ferromagnetic layer, is
\begin{equation}
\left. \frac{\partial \check f}{\partial x} \right|_{x=d}=0.
\end{equation}

Similarly to Sec.~\ref{sec:basic}, we can continue $\check f$ to the whole $x$ axis, reflecting it with respect to $x=0$
and periodically continuing from the $(-d,d)$ interval. The advantage of this trick is that the boundary conditions are
included into the equation. Similarly to Eq. (\ref{usadelM1}), we obtain
\begin{multline}
\frac{\partial^2 \check f}{\partial x^2} +\frac{\partial^2 \check f}{\partial y^2} - 2 k_\omega^2 \check f - i k_h^2
\sgn\omega \Bigl( \left\{ \hat\sigma_3, \check f \right\} \cos\alpha (y) \\
+\hat\tau_3 \left[ \hat\sigma_2, \check f \right] \sin\alpha (y) \Bigr) = -\frac{2 \check f_S}{\gamma_b}
\sum_{N=-\infty}^\infty \delta (x -2dN) ,
\end{multline}
which must be solved at all $x$ in the class of even and $2d$-periodic functions. A periodic function can be expanded
into the Fourier series:
\begin{align}
\check f(x) &= \frac 1{2d} \sum_{n=-\infty}^\infty e^{ik_n x} \check f(k_n),\qquad k_n = \frac\pi d n, \label{sum} \\
\check f(k_n) &= \int_{-d}^d e^{-ik_n x} \check f(x) dx. \notag
\end{align}
After this transformation, we reproduce Eq. (\ref{usadelM2}) with the only difference that the continuous wave vector
$k$ is substituted by discrete $k_n$.

As in Sec.~\ref{sec:basic}, the requirement that $\check f(x)$ is even, is automatically satisfied since the equation
contains only $k_n^2$, while the $2d$-periodicity is guaranteed since we consider $k_n$ defined by Eq. (\ref{sum}).

The equivalence to the previous equations allows us to directly use the results of Secs.~\ref{sec:basic}
and~\ref{sec:inf}, obtained for the infinite $d$. The rule is very simple: in the case of finite $d$, all the results of
Secs.~\ref{sec:basic} and~\ref{sec:inf} for the Fourier harmonics are valid if we substitute $k$ by $k_n$. The
real-space function can then be calculated with the help of Eq. (\ref{sum}).

For example, we consider the case of multidomain SF structure with the F layer of thickness $d$. For the case of
positive chirality, instead of Eq. (\ref{f_1Dom}) inside of the domain ($|y|<a_0$), we obtain
\begin{equation} \label{f_1Domd}
f_L (x,y) = \frac 1{2d} \sum_{k_n} e^{ik_n x} f_1(a_0) \frac{\sinh (\kappa_0 y)}{\sinh \theta_0} ,
\end{equation}
where $f_1(a_0)$ is given by Eq. (\ref{f1a0}).

The formula (\ref{f_1Domd}) can be drastically simplified in the limit when the F film is thin for the long-range
component but thick for the short-range one (i.e., $k_h \gg 1/d \gg Q, k_\omega$). In this case, the main contribution
is given by the $n=0$ harmonic, since otherwise $\kappa$ in the denominator of Eq. (\ref{f1a0}) become very large.
Therefore, Eq. (\ref{f_1Domd}) yields
\begin{multline}
f_L (x,y)= -\left( \frac{i f_S \sgn\omega}{2d \gamma_b k_h^2} \right) \frac Q{\kappa_Q \coth \theta_Q +\kappa_0 \coth
\theta_0} \\
\times \left. \frac{\sinh (\kappa_0 y)}{\sinh (\kappa_0 a_0)} \right|_{k_n=0} ,
\end{multline}
where we have used Eq. (\ref{F0}). The $x$ dependence has vanished since the F layer is thin.

\begin{figure}
 \includegraphics[width=8cm]{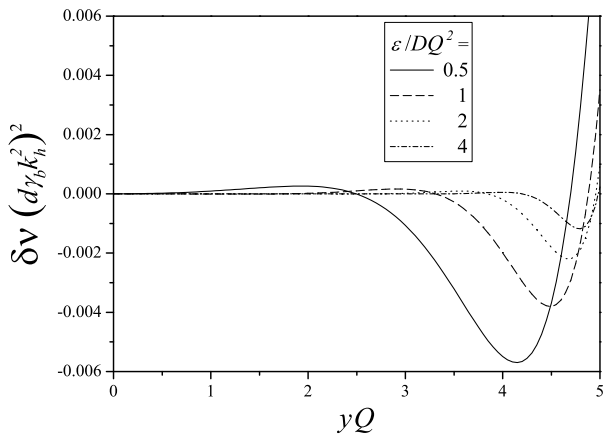}
\caption{Correction $\delta\nu(y)$ (due to the proximity effect) to the DOS at the free surface of the F layer in the
case of positive chirality. $\delta\nu$ is multiplied by a dimensionless parameter $\left( d\gamma_b k_h^2 \right)^2$,
while $y$ is normalized by $Q$. The curves are plotted at several energies $\varepsilon$ (normalized by $DQ^2$). The
width of the domains is $a_0 = 5/Q$, while the rotation of magnetization in the domain walls corresponds to $Q a_Q
=\pi$. We assume the limit $DQ^2 \ll \Delta$, which means that the width of the domain walls is larger than the
coherence length. At the boundary of applicability of our approximation, $\delta\nu(y)$ falls into experimentally
measurable range\cite{Kontos2} (see text for details).}
 \label{fig:DoS}
\end{figure}

The variation of the DOS in space, $\delta \nu (y) = - \left. \Real f_L^2 /2 \right|_{\omega\to -i\varepsilon}$, differs
drastically from the case without the LRTC: it is almost constant across the layer (no $x$ dependence) and equal to zero
in the middle of domains ($y=0$). The condensate function $f_L (y)$ decays exponentially (oscillating at the same time)
from the boundaries between domain walls and domains with characteristic length $\xi_T = \sqrt{D/2\pi T}$ (for the DOS,
$\xi_\varepsilon = \sqrt{D/\varepsilon}$); see Fig.~\ref{fig:DoS} for illustration. At the same time, the singlet
component in the case of a domain structure leads to an even dependence of $\delta \nu (y)$ with respect to the middle
of a domain. The characteristic length of the short-range components is $\xi_h = \sqrt{D/h}$. This case is realized if
$\theta_Q \to 0$.

The case of negative chirality is treated similarly, and inside of the domain ($|y|<a_0$) we obtain
\begin{equation} \label{f_1Domd_nc}
f_L (x,y) = \frac 1{2d} \sum_{k_n} e^{ik_n x} f_1(a_0) \frac{\cosh (\kappa_0 y)}{\cosh \theta_0} ,
\end{equation}
where $f_1(a_0)$ is given by Eq. (\ref{f1a0_nc}). In the limit $k_h \gg 1/d \gg Q, k_\omega$, Eq. (\ref{f_1Domd_nc})
yields
\begin{multline}
f_L (x,y)= -\left( \frac{i f_S \sgn\omega}{2d \gamma_b k_h^2} \right) \frac Q{\kappa_Q \coth \theta_Q +\kappa_0 \tanh
\theta_0} \\
\times \left. \frac{\cosh (\kappa_0 y)}{\cosh (\kappa_0 a_0)} \right|_{k_n=0} .
\end{multline}

\begin{figure}
 \includegraphics[width=8cm]{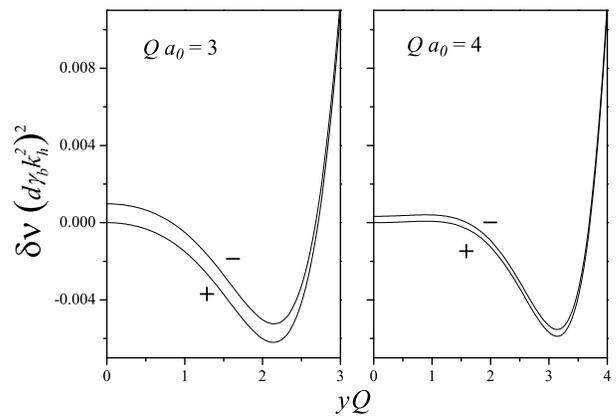}
\caption{Correction $\delta\nu(y)$ to the DOS at the free surface of the F layer: comparison of positive (denoted by
``$+$'') and negative (denoted by ``$-$'') chiralities at two widths of domains, $a_0 = 3/Q$ and $a_0 = 4/Q$. The energy
is $\varepsilon = 0.5 D Q^2$. The difference between the chiralities becomes less pronounced as $a_0$ or $\varepsilon$
increases.}
 \label{fig:DoS_pn}
\end{figure}

The resulting corrections to the DOS, $\delta\nu(y)$, in the case of positive and negative chiralities are compared in
Fig.~\ref{fig:DoS_pn}.

The dimensionless parameter $\left( d\gamma_b k_h^2 \right)^2$, which multiplies $\delta\nu(y)$ in Figs.~\ref{fig:DoS}
and~\ref{fig:DoS_pn}, is much larger than unity under our assumptions. This means that $\delta\nu(y)$ is very small.
However, at the boundary of our approximation, when the parameter is of the order of unity, we expect no qualitative
differences to appear. At the same time, in this case the DOS correction shown in Figs.~\ref{fig:DoS}
and~\ref{fig:DoS_pn} already falls into experimentally measurable range: in Ref.~\onlinecite{Kontos2}, the resolution of
the DOS measurement was of the order $0.002\div 0.003$. Experimentally, it is desirable to measure $\delta\nu$ at small
energies, since the effect grows as energy decreases, see Fig.~\ref{fig:DoS}.

The behavior of the LRTC [and the corresponding correction to the DOS $\delta\nu(y)$] inside of the domain walls is
similar to Eqs. (\ref{f_1Domd}) and (\ref{f_1Domd_nc}) with the main difference that the decay length in the $y$
direction is determined not by $\kappa_0$ but by $\kappa_Q$.

The coordinate dependence of the DOS presented in Figs.~\ref{fig:DoS} and~\ref{fig:DoS_pn} differs from the
corresponding dependence of the DOS variation caused by the singlet component. If the thickness of the F film $d$ is
less than the exchange length $\xi_h$, then the DOS variation, related to the singlet component, in a domain wider than
$\xi_h$, does not depend on $y$. In the case of the LRTC the DOS variation is a strongly coordinate-dependent quantity
since the LRTC arises near the domain walls.

\section{Conclusions} \label{sec:concl}

We have studied the long-range triplet superconductivity or LRTC in the SF multidomain structure, where magnetic domains
are separated by the N\'eel domain walls. The magnetization vector $\mathbf M$ is supposed to lie in the plane of the F
film. We show that in this case the LRTC arises at the domain walls and decays in domains over a large distance $\xi_T
=\sqrt{D/2\pi T}$ (we assume the diffusive case). The same length characterize the decay of the LRTC from the SF
interface. Although the amplitude of the LRTC is less than the amplitude of the singlet component in the F film at the
interface by $k/k_h$ times, it decays much more slowly in the F film (here the characteristic value of the wave vector
$k$ is of the order $\xi_T$ or $d^{-1}$). Therefore if the thickness of the F film $d$ is much larger than the
``exchange'' length $\xi_h =\sqrt{D/h}$, which characterizes the decay of the singlet component in the ferromagnet, only
the LRTC $f$ survives at the outer surface of the F film in a SF structure. Its spatial in-plane dependence in domains
$f(y)$ differs drastically from the corresponding dependence of the singlet component near the SF interface. If the
vector $\mathbf M(y)$ rotates in the N\'eel walls in the same direction (positive chirality), then $f$ turns to zero in
the centers of domains. This implies that the DOS variation due to the proximity effect $\delta \nu (\varepsilon ,y)
\sim f^2 (\varepsilon ,y)$ varies inside a domain turning to zero in the middle. Thus the measurements of the DOS
variation at the outer surface of the F film in SF bilayers allows one to get an information on the nature of the
condensate in the ferromagnetic films (singlet or triplet).

The effects of the LRTC are mostly pronounced at the boundaries between domains and domain walls. However, the
correction to the DOS that we find is small since we assume the SF interface of low transparency and hence a weak
proximity effect. At the same time, as we increase the interface transparency and reach the limit of applicability of
our approximation, the correction to the DOS falls into experimentally measurable range, while we expect that our
results are still qualitatively valid. Another way to increase the correction to the DOS is to measure it at smaller
energies.

The obtained results provide an insight for the Josephson effect in SFS junctions with multidomain structure in the F
layer. If the thickness of the F layer $d$ is much larger than the short exchange length $\xi_h$, the Josephson coupling
between the S layers is due to the LRTC. In this case the local critical current density $j_c (y)$ is modulated in
space, reaching its maxima at the domain walls and decaying to the centers of domains.

\begin{acknowledgments}
We are grateful to A.~A. Golubov and M.~V. Feigel'man for helpful discussions. We would like to thank SFB 491 for
financial support. Ya.V.F. was also supported by the RFBR Grants Nos. 04-02-16348 and 04-02-08159, the RF Presidential
Grant No.~MK-3811.2005.2, the Russian Science Support Foundation, the Russian Ministry of Industry, Science and
Technology, the program ``Quantum Macrophysics'' of the Russian Academy of Sciences, CRDF, and the Russian Ministry of
Education.
\end{acknowledgments}

\appendix*

\section{Calculating $A_{il}$ and $B_{il}$} \label{sec:app}

To find the coefficients $A_{il}$ and $B_{il}$, we match solutions (\ref{A}) and (\ref{B}) with the help of boundary
conditions (\ref{MC1}) and (\ref{MC2}). This yields
\begin{gather}
A_{3+} +A_{3-} +F_3 (Q) =B_{3+}+B_{3-}+F_3 (0) , \label{A1} \\
-A_{3+}+A_{3-}+F_0 (Q) =-B_{3+}+B_{3-}+F_0 (0) , \label{A2} \\
-\frac{2Q}{\kappa_+} A_{3+} +\frac{2Q}{\kappa_-} A_{3-}+A_{1L}=B_{1L} , \label{A3} \\
-\kappa_+ A_{3+} -\kappa_- A_{3-}=\kappa_+ B_{3+}+\kappa_- B_{3-} , \label{dA1} \\
\kappa_+ A_{3+} -\kappa_- A_{3-} -Q B_{1L} =-\kappa_+ B_{3+}+\kappa_- B_{3-} , \label{dA2} \\
Q \left[ A_{3+}-A_{3-}+F_0 (Q) \right] -\kappa_Q A_{1L} =\kappa_0 B_{1L} , \label{dA3}
\end{gather}
where for brevity we have denoted
\begin{equation}
\kappa_Q \equiv \kappa_L(Q),\quad \kappa_0 \equiv \kappa_L(0).
\end{equation}

From Eqs. (\ref{dA1}) and (\ref{dA2}) we find
\begin{align}
B_{3+} &= -A_{3+}+ \frac Q{2\kappa_+} B_{1L}, \\
B_{3-} &= -A_{3-}- \frac Q{2\kappa_-} B_{1L}.
\end{align}
From Eqs. (\ref{A1}) and (\ref{A2}) we find
\begin{align}
2(A_{3+}+A_{3-}) +\frac Q2 \left( \frac 1{\kappa_-} -\frac 1{\kappa_+} \right) B_{1L} &= -\delta F_{3}, \label{A+A} \\
2(-A_{3+}+A_{3-})+\frac Q2 \left( \frac 1{\kappa_-} +\frac 1{\kappa_+} \right) B_{1L} &= -\delta F_{0}, \label{-A+A}
\end{align}
where $\delta F_3 =F_3 (Q)-F_3 (0)$ and $\delta F_0 =F_0(Q)-F_0(0)$. If Eq. (\ref{Assum}) is fulfilled, on the order of
magnitude we have $\delta F_3 \propto (Q/k_h)^2 F_0$ and $\delta F_0 \propto (Q/k_h)^2 ((k^2+2k_\omega^2) /k_h^2) F_0$,
where
\begin{equation} \label{F0}
F_0 \approx -\frac{i f_S \sgn\omega}{\gamma_b k_h^2} .
\end{equation}
It follows from Eqs. (\ref{A+A}) and (\ref{-A+A}) that the coefficients $A_{3\pm}$ are smaller than $B_{1L}$ by the
parameter $Q/k_h$.

From Eq. (\ref{A3}) we find
\begin{equation}
A_{1L} \approx B_{1L}.
\end{equation}
From Eq. (\ref{dA3}) we find
\begin{equation}
Q (A_{3+}-A_{3-}) -\kappa_Q A_{1L} +Q F_0 =\kappa_0 B_{1L}.
\end{equation}
The first term here is small, therefore we finally obtain
\begin{equation} \label{A1LB1L}
A_{1L} \approx B_{1L} \approx \frac{Q F_0}{\kappa_Q +\kappa_0} .
\end{equation}

These coefficients determine the amplitude of the LRTC in the domain ($y<0$) and in the region with rotating
magnetization ($y>0$).

\end{document}